\documentclass[preprint,showpacs,preprintnumbers,amsmath,amssymb]{revtex4}

\usepackage{graphicx}
\usepackage{dcolumn}
\usepackage{bm}

\begin{document}

\preprint{APS/123-QED}

\title{Asymptotic Behavior of the Fokker-Planck Type Equation
and Overall Phase Transformation Kinetics}

\author{Yoshiyuki Saito}
 \email{ysaito@mn.waseda.ac.jp}
\affiliation{%
Department of Materials Science and Engineering, Waseda University\\
3-4-1 Okubo, Shinjuku-ku, Tokyo 169-8555, Japan
}%

\date{\today}

\begin{abstract}
This papers deals with overall phase transformation kinetics. The Fokker-Planck type equation is derived from the generalized nucleation theory proposed by Binder and Stauffer. Existence of the steady state solution is shown by a method based on the mean value theorem of differential calculus. From the analysis of asymptotic behavior of the Fokker-Planck type equation it is known that the number of clusters having the critical size increases with time in the case of constant driving force. On the basis of the present study on overall phase transformation kinetics a simple method for analyzing experimental phase transformation curves was proposed. 
\end{abstract}

\maketitle
\section{Introduction}
Overall phase transformation kinetics can easily be detected. A number of techniques allow in situ detection. Usually the results are plotted as changes in physical property, such as the dilatation and the electrical resistance, versus time. Some of those observation can be transformed into the volume fraction of newly precipitated phase versus time.

The progress of isothermal phase transformation kinetics can be conveniently representing by plotting the fraction transformed as a function of time, phase transformation curve. 
The Avrami type of equation \cite{av} has been utilized for the analysis of given phase transformation curve. The fraction transformed $f(t)$  is described by a equation with the form
\begin{equation}
\label{eq:ph1}
f(t)=1-\exp(-Ct^n)
\end{equation}
where $C$ and $n$ are constants which are dependent on temperature and a composition of material. Eq.(\ref{eq:ph1}) is reduced to
\begin{equation}
\ln\left[\ln\frac{1}{1-f(t)}\right]=C+n \ln t
\end{equation}
We assume linear relationship between $\ln t$ and $\ln[\ln(1/(1-f(t)))]$ and then determine a numerical exponent $n$ by the slope of the line. The value of $n$ varies from 1 to 4 and it depends on dominating mechanism of phase transformation.  However applicability of the Avrami equation is very limited. 

The overall phase transformation kinetics is known to be descibed by the Kolmogorov-Johnson-Mehl-Avrami(KJMA) equation \cite{av}$^-$\cite{chr}(see Appnedix A).
\begin{equation}
\label{eq:ph3}
f(t)=1-\exp\left[-\int_0^t J(t')V(t,t')dt' \right]
\end{equation}
where $f(t)$ is the fraction transformed, $J(t)$ is the nucleation rate at time $t$ and $V(t,t')$ is the volume at $t$ of a stable nucleus formed at time $t'$.
The Avrami equation is the special case of the KJMA equation in which we assume
\begin{eqnarray}
J(t)&=&C_0\delta (t)
\nonumber \\
V(t,t')&=&C_1t^n
\nonumber
\end{eqnarray}
where $\delta (x)$ is the Dirac delta function and $C_0$ and $C_1$ are constants. In the above case, we assume all nuclei are formed at time zero and the stable nuclei grow with a constant speed.

In this paper we will deal with a simple model for overall phase transformation kinetics. First the Fokker-Planck type equation will be derived from the generalized nucleation theory proposed by Binder and Stauffer\cite{bs}. Asymptotic behavior of the Fokker-Planck type equation will be investigated by using the general mean value theorem of differential calculus. On the basis of the KJMA equation a simple method for analyzing experimental phase transformation curve will be
 described.
\section{Asymptotic behavior of the Fokker-Planck type equation}
\subsection{Cluster dynamics approach toward generalized
nucleation theory}
     
Binder and Stauffer\cite{bs,bs1} proposed a dynamical theory of cluster formation in which dissociation-coagulation mechanisms are caused by "cluster reactions". It deals with a wide range of cluster sizes, which include both small and large clusters. This theory is considered to be a natural extension of the Becker-D$\rm{\ddot{o}}$ring theory\cite{bd}. Binder and Stauffer considered the time evolution of the cluster size distribution $n(l,t)$ in terms of a cluster coagulation or a cluster splitting mechanism. If we ignore cluster diffusion, we can write an equation for evolution of the number of clusters of size l, $n(l,t)$, 
\begin{eqnarray}
\frac{dn(l,t)}{dt} &=&\sum_{l'=1}^\infty S_{l+l'\;l} n(l+l',t) - \frac{1}{2}\sum_{l'=1}^{l-1}S_{l\; l'}n(l,t) \nonumber \\ &+&\frac{1}{2}\sum_{l'=1}^{l-1}C_{l-l'\; l'}n(l',t) n(l-l',t)
\nonumber \\
&-&\sum_{l'=1}^\infty C_{l\; l'} n(l,t)n(l',t)   
\label{eq:cd1}
\end{eqnarray} 
The first term on the right hand size of Eq.(\ref{eq:cd1}) accounts for an increase in the number of $l$-clusters due to the splitting reaction $[l+l']\to [l]+[l']$. This is assumed to be proportional to the number of clusters of size $(l+l') $ which are present, where $S_{l+l' l'} $ is the rate constant. The second term represents a decrease in  the number of $l$-clusters caused by  the splitting reaction $[l]\to [l-l']+[l']$. The factor 1/2 in the second and the third terms accounts for overcounting pairs in the summation. The third term represents  an increase in the number of $l$ clusters by coagulation of clusters $[l-l']+[l']\to [l]$. This is a reverse process of that represented by the second term where $C_{l-l' l'} $ is the rate constant. The fourth term represent a decrease of $l$-clusters by coagulation $[l]+[l']\to [l+1]$.                                              
The next step is to assume the detailed balance condition. The rates $S$ and $C$ can be replaced by a single reaction rate as 
\begin{equation}                                         
S_{l+l'\;l'} n_0(l+l') =C_{l\;l'}n_0(l) n_0(l') =W(l,l')         
\label{eq:cd2}
\end{equation}
where $n_0(l)$ is the equilibrium cluster distribution function and given by
\begin{equation}
n_0(l)=n_1\exp\left[-\frac{\epsilon(l)}{k_BT}\right]
\label{eq:cd2a}
\end{equation}
with
\begin{equation}
\epsilon(l)=-\delta\mu l-\gamma l^{2/3}
\label{eq:cd2b}
\end{equation}
A function $\epsilon (l)$ in Eq.(\ref{eq:cd2a}) is the free energy of formation of a cluster of size $l$ which contains a bulk part of magnitude $\delta\mu$ and a surface term with a coefficient $\gamma $ that is proportional to the surface energy.
Substitution of Eq.(\ref{eq:cd2}) into Eq. (\ref{eq:cd1}) yields
\begin{eqnarray}
\frac{dn(l,t)}{dt} &=&\sum_{l'=1}^\infty W(l,l')\frac{ n(l+l',t)}{n_0(l+l')} 
\nonumber \\
&-&\frac{1}{2}\sum_{l'=1}^{l-1}W(l-l',l')\frac{n(l,t)}{n_0(l)}
 \nonumber \\
&+&\frac{1}{2}\sum_{l'=1}^{l-1}W(l-l',l')\frac{n(l',t) n(l-l',t)}{n_0(l')n_0(l-l')}
\nonumber \\ &-&\sum_{l'=1}^\infty W(l,l')\frac{n(l,t)n(l',t)}{n_0(l)n_0(l')}   
\label{eq:cd3A}
\end{eqnarray} 
Let us consider the case $l>>l'$. If we assume that dissociation-coagulation mechanisms are controlled by small clusters such that $l'\le[l/2]$, Equation (\ref{eq:cd3A}) is rewritten as
\begin{eqnarray}
\frac{dn(l,t)}{dt} &=&\sum_{l'=1}^{[l/2]}W(l,l')\frac{ n(l+l',t)}{n_0(l+l')} 
\nonumber \\
&-&\sum_{l'=1}^{[l/2]}W(l-l',l')\frac{n(l,t)}{n_0(l)}
 \nonumber \\ &+& \sum_{l'=1}^{[l/2]}W(l-l',l')\frac{n(l',t) n(l-l',t)}{n_0(l')n_0(l-l')} \nonumber \\
&-&\sum_{l'=1}^{[l/2]}W(l,l')\frac{n(l,t)n(l',t)}{n_0(l)n_0(l')}   
\label{eq:cd3}
\end{eqnarray}
We assume that the concentration of small clusters are that of the supersaturated one-phase equilibrium.
The deviation from equilibrium of $n(l',t)/n_0(l')$ is neglected in the nonlinear terms, i.e. $n(l',t)/n_0(l')=1$. Equation (\ref{eq:cd3}) is simplified as:
\begin{eqnarray}
\frac{dn(l,t)}{dt} &=&\sum_{l'=1}^{[l/2]}\biggl[W(l,l')\left[\frac{ n(l+l',t)}{n_0(l+l')} - \frac{n(l,t)}{n_0(l)}\right] 
\nonumber \\
&+&[W(l-l',l')-w(l,l')]\left[\frac{n(l-l',t)}{n_0(l-l')}-\frac{n(l,t)}{n_0(l)}\right] \biggr]\nonumber \\
\label{eq:cd3a}
\end{eqnarray}
By expanding the terms $n(l+l',t)/n_0(l+l')$ and $W(l-l',l')$ about $l$, we obtain
\begin{equation}
\frac{n(l\pm l',t)}{n_0(l\pm l',t)}=\frac{n(l,t)}{n_0(l)}\pm
 l'\frac{\partial}{\partial l}\left[\frac{n(l,t)}{n_0(l)}\right]+\frac{(l')^2}{2}\frac{\partial ^2}{\partial ^2l}\left[\frac{n(l,t)}{n_0(l)}\right]
\label{eq:cd3b}
\end{equation}
\begin{equation}
W(l-l',l')=W(l,l')-l'\frac{\partial W(l,l')}{\partial l}
\label{eq:cd3c}
\end{equation}
Substitution of Eqs.(\ref{eq:cd3b}) and (\ref{eq:cd3c}) into Eq.(\ref{eq:cd3a}) yields a Fokker-Planck type equation
\begin{eqnarray}
\frac{\partial n(l,t)}{\partial t}&=&\frac{\partial }{\partial l}\left[a(l)n_0(l)\frac{\partial}{\partial l}\left[\frac{n(l,t)}{n_0(l)}\right]\right]
\label{eq:cd4m}
\end{eqnarray} 
where we have defined a cluster reaction rate $a(l)$ in terms of $W(l,l')$ describing reactions $[l-l']+[l']\rightleftharpoons [l]$.
\begin{equation} 
a(l)=\frac{1}{n_0(l)}\sum_{l'=1}^{[l/2]}(l')^2 W(l,l') 
\end{equation}
Thus the Fokker-Planck type equation can be derived in the general case that we allow growth and shrinking not only by evaporation/condensation of monomers but also by small clusters of size $l'$. This indicates that we have formally the same expressions of steady state cluster concentration and time-dependent nucleation rate as those  given by the Becker-D${\rm \ddot{o}}$ring theory.

If we consider the contributions of larger clusters, i.e. $l'>[l/2]$, which is neglected the above treatment, Eq.(\ref{eq:cd4m}) is rewritten as
\begin{eqnarray}
\frac{\partial n(l,t)}{\partial t}&=&\frac{\partial }{\partial l}\left[a(l)n_0(l)\frac{\partial}{\partial l}\left[\frac{n(l,t)}{n_0(l)}\right]\right]
\nonumber \\
&+&\frac{1}{2}\int_{l_c}^lW(l-l',l')
\frac{n(l,t)n(l-l',t)}{n_0(l)n_0(l-l')}dl' \nonumber \\
&-&\int_{l_c}^lW(l,l')\frac{n(l,t)n(l',t)}{n_0(l)n_0(l')}dl'
\label{eq:cd4}
\end{eqnarray} 

The first term in Eq.(\ref{eq:cd4}) is the Fokker-Planck type equation. The last two terms corresponds to standard coagulation equation called Smoluchowski equation. Thus Eq.(\ref{eq:cd4}) is known to be a combination of two classic equations.

\subsection{Steady state solution of the Fokker-Planck type equation}
\begin{figure}
\includegraphics{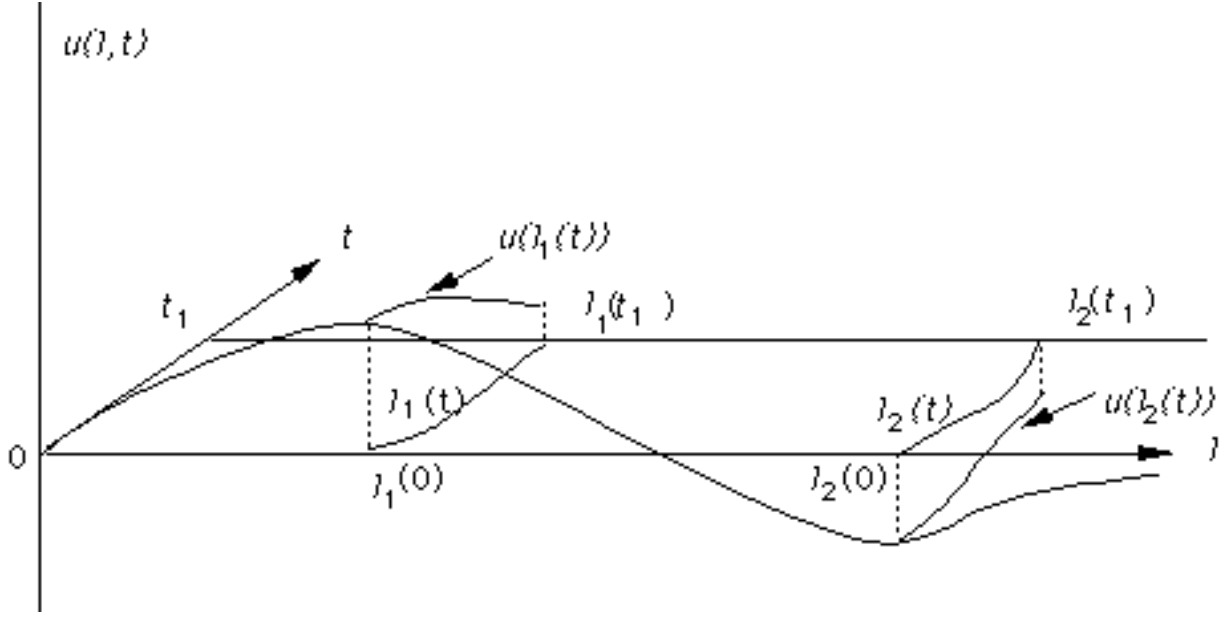}
\caption{\label{fig:fig_1} Schematic illustration of trajectories of a peak top  and a bottom of a valley of a function $u(l,t)$}
\end{figure}

First, we will show existence of a steady state solution of the Fokker-Planck
type equation. Equation (\ref{eq:cd4m}) is rewritten by introducing a 
 new variable
\begin{equation}
u(l,t)=\frac{n(l,t)-n_s(l)}{n_0(l)}
\end{equation}
as
\begin{equation}
\label{eq:nc99}
n_0(l)\frac{\partial u}{\partial t}=\frac{\partial }{\partial l}\left[a(l)n_0(l)\frac{\partial u}{\partial l}\right]
\end{equation}
where $n_s(l)$ is a time independent distribution function.
The boundary conditions  for Eq(\ref{eq:nc99}) are
\begin{equation}
u(l,t)  =  0, \quad l\rightarrow 0,  \hspace{1cm}
u(l,t)  =  0, \quad l\rightarrow\infty
\end{equation}
Unless $u(l,t)\equiv 0$, a solution $u(l,t)$ of Eq. (\ref{eq:nc99}) satisfies the above boundary condition has at least one positive maximum value and one negative minimal value(see Fig.\ref{fig:fig_1}).

Consider temporal behavior of a solution along peak top position of $u$ which satisfies $\partial u/\partial l=0$.
For $l=l_1$ 
\begin{equation}
\frac{\partial u}{\partial l } =0, \hspace{1cm}
\frac{\partial ^2u}{\partial l^2 } <0 
\end{equation}
A function $F(t,l,\partial u/\partial l,\partial ^2u/\partial l^2,u)$ is defined as
\begin{equation}
F\left(t,l,\frac{\partial u}{\partial l},\frac{\partial ^2u}{\partial l^2},u\right) \equiv 
\frac{\partial }{\partial l}\left[a(l)n_0(l)\frac{\partial u}{\partial l}\right]
\end{equation}
This function satisfies the following condition:
\begin{equation}
F(t,l_1,0,0,u)=0
\end{equation}

We utilize a fundamental relation between the difference quotient and derivative known as the mean value theorem of differential calculus \cite{couran}.
Let $f(x)$ is a continuous function on the closed interval $[a,b]$. We assume that the derivative exists everywhere in the closed interval$[a,b]$. There exits an intermediate value $\xi$ in the interval such that 
\begin{equation}
\frac{f(b)-f(a)}{b-a}=f'(\xi ) 
\end{equation}
This statement is called the mean value theorem of differential calculus.
From this ordinary mean value theory we can derive the general mean value theorem of differential calculus. We assume $u(x)$  is monotonic in the interval $[x,y]$. The compound function $f(u(x))$ for $x$ is defined for $u(x)$ in the interval $[u(x),u(y)]$.  We find that there exist a function $f'(\xi)$ for an intermediate value $\xi $ in the open interval $(u(x), u(y))$ such that
\begin{equation}
f(u(y))-f(u(x))=f'(\xi)[u(y)-u(x)] 
\end{equation}

We obtain the following equation for an intermediate value $\zeta$ in the open interval $((\partial ^2u/\partial l^2), 0)$
\begin{eqnarray}
n_0\left(\frac{du}{dt}\right)&=&F\left(t,l_1,\frac{\partial u
}{\partial l},\frac{\partial ^2u
}{\partial l^2},u\right)-
F(t,l_1,0,0,u) \nonumber \\
&=&\left(\frac{\partial ^2u}{\partial l^2}\right)\frac{\partial F(t,l_1,0,\zeta  , u)}{\partial  (\partial ^2u/\partial l^2)}<0
\end{eqnarray}
This equation indicates that the function $u$ along the peak top position  decreases with time and approaches the constant value$c_1$. By repeating the above procedure for a solution along the valley bottom position, it is known that the negative minimum value increases with time and approaches the constant value $c_2$.

\begin{figure}
\includegraphics{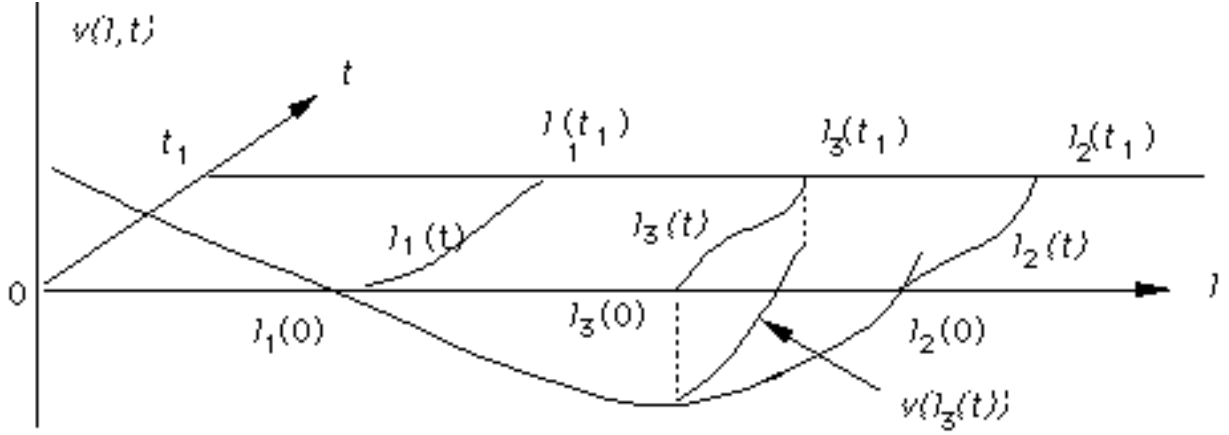}
\caption{\label{fig:fig_2} Schematic illustration of a trajectory of a bottom of a valley of a function $v(l,t)$}
\end{figure}

As the absolute values of $u$ along peak top and valley bottom positions decreases with time, the function $u$ is considered to be time independent at longer times. We will show that the constant value $c_1$ is equal to $c_2$. By differentiating the both members of Eq.(\ref{eq:nc99}) with respect to $l$ we obtain
\begin{eqnarray}
a(l)n_0(l)\frac{\partial u}{\partial l \partial t}&=&a(l)\frac{\partial ^2}{\partial l^2}\left[a(l)n_0(l)\frac{\partial u}{\partial l}\right] \nonumber \\
&-&a(l)\frac{\partial}{\partial l}[\ln n_0(l)]\frac{\partial}{\partial l}
\left[a(l)n_0(l)\frac{\partial u}{\partial l}\right]
\label{eq:ncm1}
\end{eqnarray}
By introducing a new variable $v$, 
\begin{eqnarray}
v=a(l)n_0(l)\frac{\partial u}{\partial l}
\nonumber
\end{eqnarray}
Eq.(\ref{eq:ncm1}) is rewritten as
\begin{equation}
\frac{\partial v}{\partial t}=a(l)\frac{\partial ^2v}{\partial l^2}-a(l)\frac{\partial}{\partial l}[a(l)n_0(l)]\frac{\partial v}{\partial l}
\label{eq:ncm2}
\end{equation}
$v$ is negative at any point within the open distance $(l_1,l_2)$ and $v(l_1)=v(l_2)=0$(see Fig.\ref{fig:fig_2}).

Let us consider the trajectory of $v(l,t)$ along the bottom of the valley, $l_3(t)$ 
where
\begin{equation}
\frac{\partial v(l,t)}{\partial l}=0, \quad \frac{\partial^2v(l,t)}{\partial l^2}>0
\label{eq:ncm3}
\end{equation}
From Eqs.(\ref{eq:ncm2}), (\ref{eq:ncm3}), it follows that
\begin{equation}
\frac{dv(l,t)}{dt}>0
\label{eq:ncm4}
\end{equation}
This indicates that the value of $v$ becomes 0 at longer times. Then it is known that the value of $u$ approaches 0 and we have the steady state solution.

Becker and D${\rm \ddot{o}}$ring used the following choice of boundary conditions to
obtain time-independent solution :
\begin{equation}
  \frac{n_s(l)}{n_0(l)}\rightarrow  0 ,\quad  if \quad
  l\rightarrow \infty .
\label{eq:nnc11}
\end{equation}
\begin{equation}
  \frac{n_s(l)}{n_0(l)}\rightarrow  1 ,\quad  if \quad l
  \rightarrow 0 .
\label{eq:nnc12}
\end{equation}
The steady state solution of Eq.(\ref{eq:cd4m}) is
\begin{equation}
\label{eq:nc10}
J_s=\frac{1}{\int_0^\infty \frac{{\rm{d}}l}{a(l)n_0(l)}}~~.
\end{equation}
The range of integration in Eq.(\ref{eq:nc10}) include the peak
at $l=l_c$   where $\epsilon (l)$ is maximum and the integral
can be evaluated by the saddle point approximation. 
The steady state nucleation rate, $J_s$, can be written as follows:
\begin{equation}
\label{eq:nc12}
J_s =a(l_c)\left(\frac{-\partial ^2\epsilon(l)/
    \partial l^2|_{l=l_c}}{2\pi k_{\rm{B}}T}\right)^{1/2}n_0(l_c)
= J_0 \exp \left[-\frac{\epsilon(l_c)}{k_{\rm{B}}T}\right] . 
\end{equation}
where $J_0$ is the so called nucleation prefactor. 
 The quantity $\epsilon(l_c)$ is an activation energy 
(the energy for forming critical nucleus). The factor 
$ (-\partial ^2\epsilon(l)/\partial l^2|_{l=l_c}/2\pi
k_{\rm{B}}T)^{1/2} $ in the right hand side of Eq.(\ref{eq:nc12}) is
known as  the  Zeldvich Factor, $Z$.

The above mentioned steady state nucleation rate does not provide any information on the momentary cluster distribution nor on the nucleation rate prior to reaching the steady state condition, i.e time dependent nucleation rate. It was found \cite{sa} that the nucleation rate, $J(t)$, approaches the steady state
\begin{equation}   
J(t)=J_s [1-\exp(-At)]
\label{eq:tdn1}
\end{equation}
where $6.3a(l_c)Z^2\le A \le 12.0 a(l_c)Z^2$. This result agrees with numerical calculation by Kanne-Dannetschek and Stauffer \cite{kds} where $A=6.4a(l_c)Z^2$ \subsection{Nucleation kinetics in materials with a time-independent driving force}
As shown previous section, the nucleation rate approaches a constant value, steady state nucleation rate, because the driving force of nucleation in an infinite system is time dependent. Here let us consider the case that the driving force is constant during phase transformation. 

\begin{figure}
\includegraphics{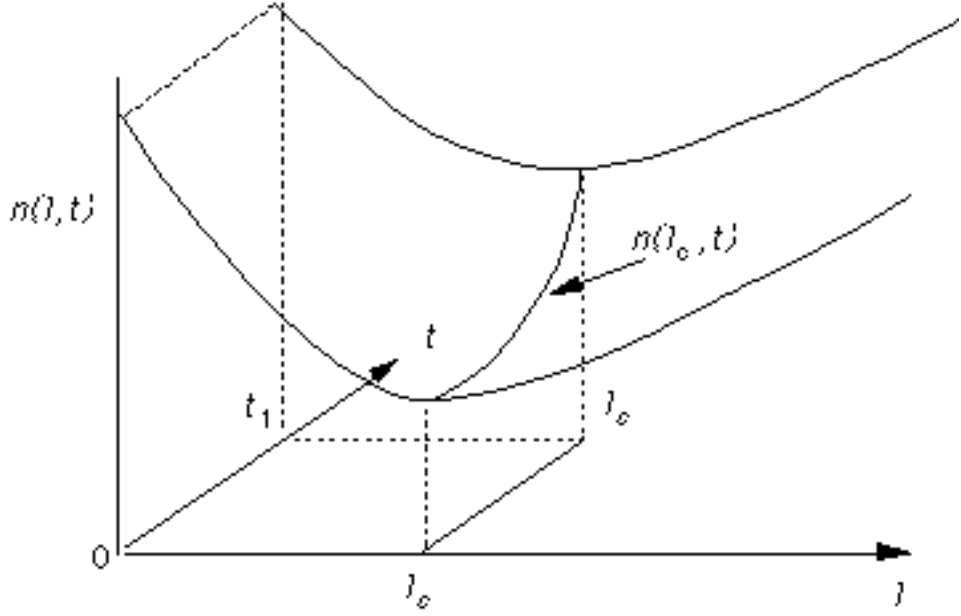}
\caption{\label{fig:fig_3} Schematic illustration of a trajectory of a bottom of a valley  of the cluster size distribution fuction $n(l,t)$ in the case of constant driving force}
\end{figure}

The Fokker-Planck type equation(\ref{eq:cd4}) is rewritten as
\begin{eqnarray}
\frac{\partial n(l,t)}{\partial t}&=&a(l)\frac{\partial^2n(l,t)}{\partial l^2}+
\left[\frac{\partial a(l)}{\partial l}+\frac{a(l)}{k_BT}\frac{\partial \epsilon(l)}{\partial l}\right]\frac{\partial n(l,t)}{\partial l} \nonumber \\
&+&\frac{1}{k_BT}\left(\frac{\partial a(l)}{\partial l}+a(l)\frac{\partial ^2 \epsilon(l)}{\partial l^2}\right)n(l,t)
\end{eqnarray}  
If the driving force of nucleation is constant the value of $\epsilon(l)$ becomes maximum at $l=l_c$. then we have
\begin{equation}
\frac{\partial \epsilon(l_c)}{\partial l}=0, \quad  \frac{\partial n_0(l_c)}{\partial l}=0, 
\end{equation}
From the assumption that the driving force is time-independent, it follows that the minimum value of $n(l,t)$  is given at $l=l_c$.
\begin{equation}
\frac{\partial n(l_c,t)}{\partial l}=0, \quad 
\frac{\partial^2n(l_c,t)}{\partial l^2}>0
\end{equation}         
From Eq.(\ref{eq:cd2b}) it follows furthermore that 
\begin{equation}
\frac{\partial^2\epsilon(l)}{\partial l^2} \rightarrow 0, \quad
l \rightarrow \infty
\end{equation}
Because very large clusters do not exist
\begin{equation}
n(l,t)\rightarrow 0, \quad l\rightarrow \infty
\end{equation}
we obtain
\begin{equation}
\frac{\partial^2n(l,t)}{\partial l^2} \rightarrow 0, \quad
l \rightarrow \infty
\end{equation}
Now we consider behavior of $n(l,t)$ along the trajectory peak top position of $n(l,t)$, $l=l_c(t)$. A fuction $G(t,l,n,\partial n/\partial l, \partial^2n/\partial l^2)$ is difined as
\begin{equation}
G\left(t,l,n,\frac{\partial n}{\partial l}, \frac{\partial^2n}{\partial l^2}\right)\equiv \frac{\partial}{\partial l}\left[a(l)n_0(l)\frac{\partial}{\partial l}\left[\frac{n(l,t}{n_0(l)}\right]\right]
\end{equation}
The fuction $G$ satisfies
\begin{equation}
G(t,l_c,n(l_c,t),0,0)=0 
\end{equation}
Applying the mean value theorem of differential calculus for compound functions to the function $G$, we obtain the following equation for an intermediate value $\zeta$ in the open interval $(0, \partial^2n/\partial l^2)$
\begin{eqnarray}
\frac{dn(l,t)}{dt}&=&G\left(t,l_c,n(l_c,t),\frac{\partial n(l_c,t)}{\partial l}, \frac{\partial^2n(l_c,t)}{\partial l^2}\right) \nonumber \\
&-&G(t,l_c,n(l_c,t),0,0) \nonumber \\
&=&\frac{\partial^2n(l_c,t)}{\partial l^2}\frac{\partial G(t,l_c,n(l_c,t),0,\zeta)}{\partial(\partial^2/\partial l^2)}\nonumber \\
&>&0
\label{eq:mvt1}
\end{eqnarray}
From Eq.(\ref{eq:mvt1}) it follows that the number of stable clusters increases with time in the case that the driving force is time-independent.(see Fig.\ref{fig:fig_3}).

\section{Overall Phase Transformation Kinetics}
Here we propose a method for analyzing given phase transformation curve with use of the KJMA equation. This method is based on the following assumptions.

\noindent (1)The time dependent nucleation rate, $J(t)$, is described by
\begin{eqnarray}
J(t)=J_s[1-\exp(-At)]
\nonumber
\end{eqnarray}
where $J_s$ and $A$ are constants.

\noindent (2) The volume at $t$ of stable nuclei formed at time $t'$, $V(t,t')$ is a function of $t-t'$, that is $V(t,t')=V(t-t')$.

\noindent As shown in the previous section, the assumption (1) is reasonable one and the assumption (2) is acceptable unless the supersaturation of metastable solid solution changes rapidly.

\begin{figure}
\includegraphics[width=15cm]{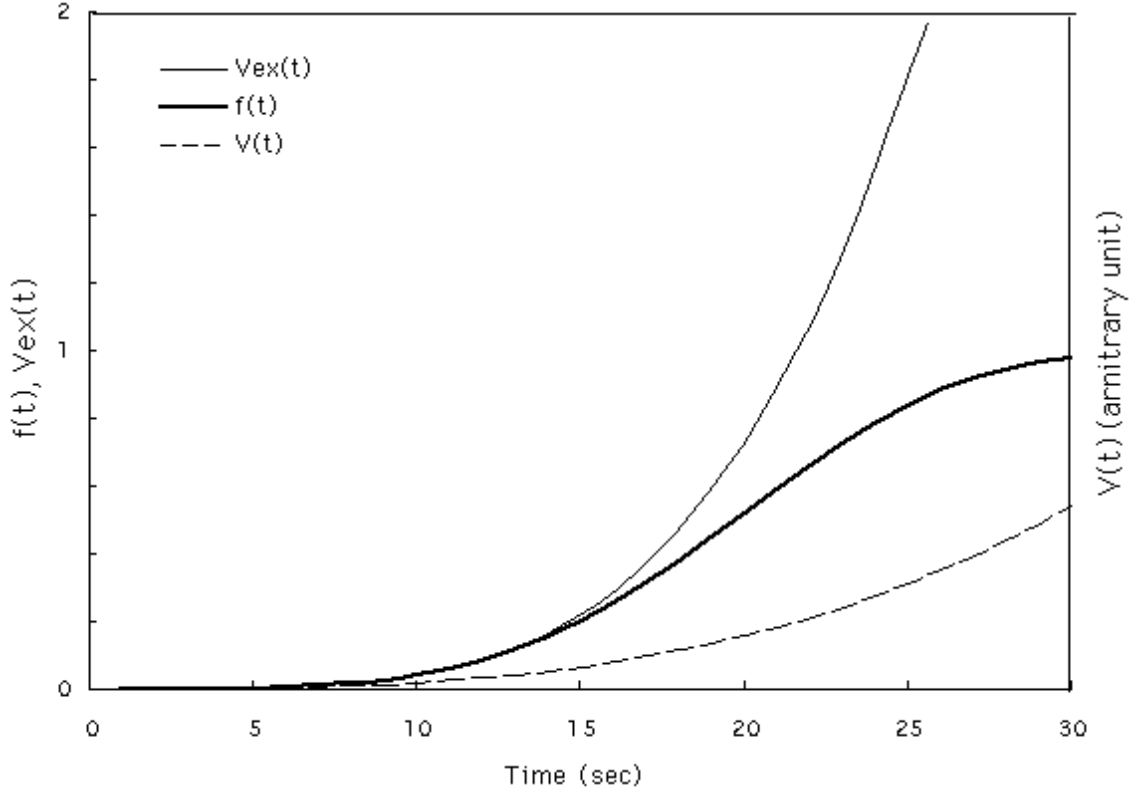}
\caption{\label{fig:fig_4} Estimation of the volume of stable nucleus $V(t)$ by a phase transformation curve, $f(t)$ versus time}
\end{figure}

First we determine the variation of the extended volume, $V_{ex}=\int_0^t J(t')$$V(t,t')dt'$, with time by given phase transformation curve.
\begin{equation}
V_{ex}=\ln\left[\frac{1}{1-f(t)}\right]
\end{equation}
From the assumptions (1),(2), it follows that
\begin{equation}
V_{ex}(t)=J_s\left[\int_0^t V(t-t') [1-\exp(-At')]dt' \right]
\end{equation}
By perfoming change of variable as $t''=t-t'$ we obtain
\begin{eqnarray}
V_{ex}(t)&=&J_s\left[\int_0^t V(t'')dt''\right. \nonumber \\
&-& \left.\exp(-At)\int_0^tV(t'')\exp(At'')dt'' \right]
\label{eq:ph6}
\end{eqnarray}
By differentiating both members of Eq.(\ref{eq:ph6}) with respect to $t$ and with use of the relation $V(0)=0$ we have
\begin{equation}
\frac{\exp(At)}{AJ_s}\frac{dV_{ex}}{dt}=\int_0^tV(t'')\exp(At'')dt''
\label{eq:ph7}
\end{equation}
Further differentiation of both members of Eq.(\ref{eq:ph7}) with respect to $t$ yields
\begin{equation}
V(t)=\frac{1}{J_sA}\left[A\frac{dV_{ex}}{dt}+\frac{d^2V_{ex}}{dt^2}\right]
\label{eq:ph8}
\end{equation}
where the constant $J_s$ is an adjustable parameter which can be determined by given phase transformation curve and the parameter $A$ can be estimated by the numerical calculation of the Fokker-Planck type equation\cite{kds} to be $6.4 a(l_c) Z^2$. Thus we can determine the change in the volume of stable nucleus with time by given phase transformation curve as shown in Fig.\ref{fig:fig_4}.

\begin{figure}
\includegraphics[width=15cm]{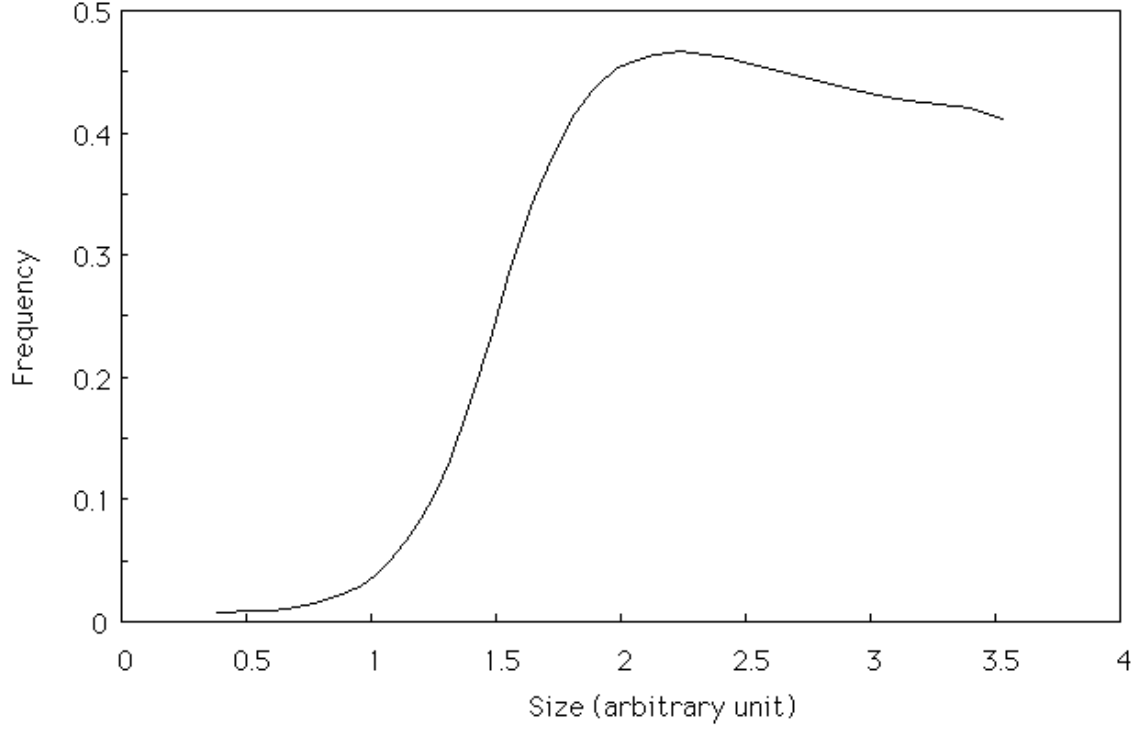}
\caption{\label{fig:fig_5} The size distribution function estimated by the  phase transformation curve shown in Fig. \ref{fig:fig_4}}
\end{figure}

Next we consider a method to estimate the size distribution function by given phase transformation curve. Let $\Delta f(t_0)$ to be a volume of nuclei formed at a interval of time between $t_0$ and $t_0+\Delta t$. The contribution of nuclei formed after a time $t''$ to the fraction transformed at time $t$, $f'(t,t'')$, is given by
\begin{equation}
\label{eq:ph9}
f'(t,t'')=1-\exp\left[-\int_{t''}^t J(t')V(t,t')dt' \right]
\end{equation}
Then we obtain $\Delta f(t_0)$ as
\begin{eqnarray}
\Delta f(t_0)&=&f'(t,t_0)-f'(t,t_0+\Delta t)\nonumber\\
    &=&\exp\left[-J_s\int_{t_0+\Delta t}^tV(t-t')[1-\exp(-At')]dt'\right]
\nonumber \\
     &-&\exp\left[-J_s\int_{t_0}^tV(t-t')[1-\exp(-At')]dt' \right] 
\end{eqnarray}
As $\Delta t \to 0$ 
\begin{eqnarray}
\lim_{\Delta t \to 0}\frac{\Delta f(t_0)}{\Delta t} &=& \frac{d}{dt_0}\Bigg[\exp\Bigg[-J_s\int_{t_0}^tV(t-t')
\nonumber \\
&\cdot&[1-\exp(-At')]dt' \Bigg] \Bigg]
\label{eq:ph12}
\end{eqnarray}
Equation (\ref{eq:ph12}) yields
\begin{eqnarray}
\lim_{\Delta t \to 0} \frac{\Delta f(t_0)}{\Delta t} 
&=&J_sV(t-t_0)[1-\exp(-At_0)] \nonumber \\
&\exp &\left[-J_s\int_{t_0}^tV(t-t')[1-\exp(-At')]dt' \right] 
\nonumber \\
\label{eq:ph12a}
\end{eqnarray}
At a interval of time between $t_0$ and $t_0+\Delta t$, $J_s[1-\exp(-At_0)]$ nuclei formed per unit volume. As the fraction transformed at the time $t_0$ is $f(t_0)$, the number of nuclei is given by $J_s[1-f(t_0)][1-\exp(-At_0)]$. Hence the average volume at time $t$ of nuclei formed at a interval of time between $t_0$ and $t_0+\Delta t$ is given by
\begin{equation}
V_{av}=\frac{V(t-t_0)}{1-f(t_0)}
\exp\left[-J_s\int_{t_0}^tV(t-t')[1-\exp(-At')]dt' \right] 
\label{eq:ph13}
\end{equation}
The fuction $V(t-t')$ in Eq.(\ref{eq:ph13}) can be determined by given phase transformation curve. With use of Eq.(\ref{eq:ph13}) we can estimate the size distribution function of nuclei. Figure \ref{fig:fig_5} shows the size distribution of nuclei estimated by the phase transformation curve in Fig. \ref{fig:fig_4}.

\section{Conclusion}
In this paper we attempted to analyze asymptotic behavior of the Fokker-Planck type equation which is derived from the generalized nucleation theory by Binder and Stauffer. From the analysis it follows that the number of cluster having the critical size increases with time if the driving force is constant during phase transformation. However, the chemical driving force which mainly controls diffusional phase transformation decreases with time. Therefore, the number of stable clusters increases with time in the case that an external driving force is applied to compensate the decrease in the chemical driving force. One of the most effective way is to give the large amount of plastic deformation to metastable phase at higher supersaturation. This may be helpful to develop a new method for producing structural materials. 

Next on the basis of a model for overall phase transformation kinetics, we propose a simple method for analyzing experimental phase transformation curves. This method is based on the Kolmogorov-Johnson-Mehl-Avrami equation. Time-dependent nucleation rate is determined by the analysis of asymptotic behavior of nucleation. A function for describing the diffusional growth can be estimated by the phase transformation curve. Furthermore the size distribution function of newly precipitated phase can be predicted with use of the present analysis.

\appendix

\section{Generalized Kolmogorov-Johnson-Mehl-Avrami Equation}

\begin{figure}
\includegraphics{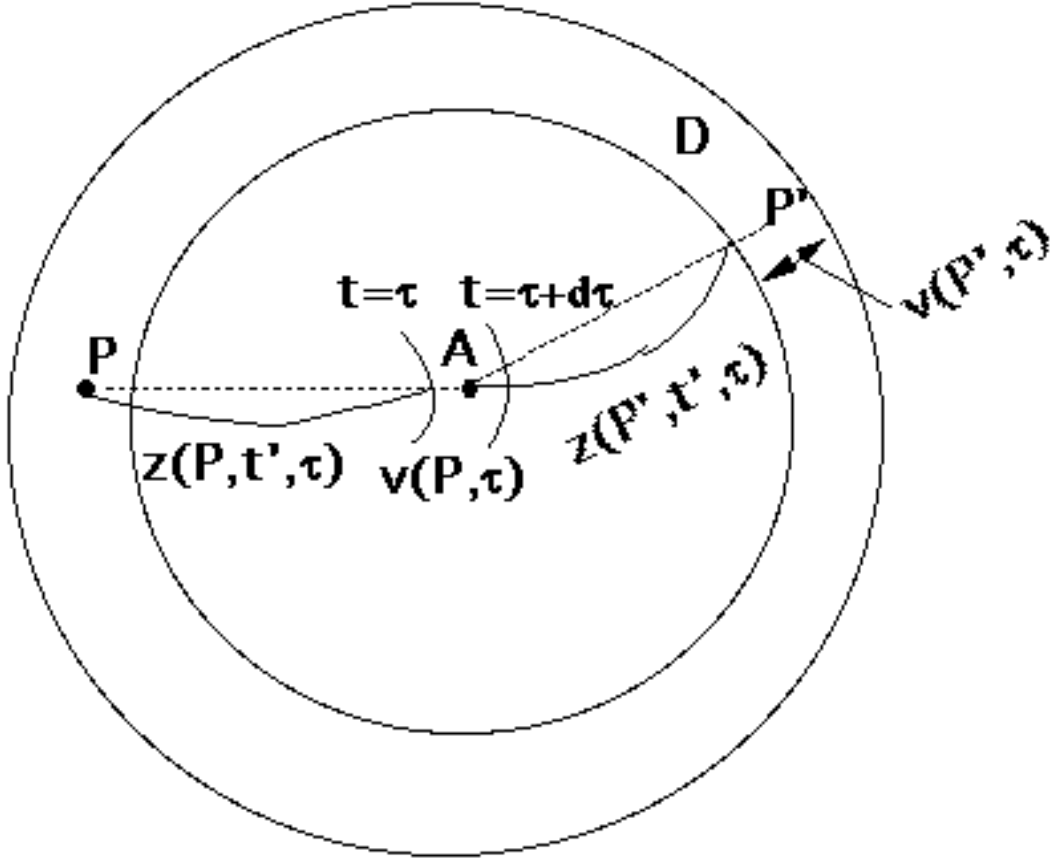}
\caption{\label{fig:fig_A1} Condition for a point A reached by a droplet at moment $t'$ within the period between $\tau$ and $\tau+d\tau$}
\end{figure}

We derive a formula for the time evolution of the volume fraction of the transformed phase at $t$, $f(t)$\cite{chr}. The volume fraction of the metastable phase is given by
\begin{equation}
g(t)=1-f(t)   
\end{equation}
Take a point A at random. We calculate the rate of the fraction transformed at the point A within the period $ \tau $ to $\tau +d\tau $ 
\begin{equation}
 df(\tau )=-dg(\tau )=-[g(\tau +d\tau )-g(\tau )] 
\end{equation}
Let $v(P,t)$ be the growth rate at time t of a droplet  which nucleated at a point P at a time $t'$. The distance between the point P and  a point Q, which is the intersection of the line AP and the surface of the droplet(see Fig.\ref{fig:fig_A1}), at time $t"$ is given by 

\begin{equation}
\overline{PQ}=Z(P,t",t')=\int _{t'}^{t''}v(P,t)dt
\end{equation}
The surface of a droplet formed at a point P' at moment $t'$ reaches the point concerned within the  period $\tau$ to $\tau+d\tau $ only if the distance between A and P' satisfies the following condition

\begin{equation}
\int_{t'}^{\tau} v(P',t)dt<\overline{P'A}<\int_{t'}^{\tau +d\tau}v(P',t)dt
\end{equation}
Thus the point A is reached by droplets formed at time $t'$  within the period between $\tau $ and 
$\tau +d\tau $ only if nucleation occurs points within a closed domain of volume 
\begin{equation}
V_A=\int_D \int Z(P',\tau ,t')v(P',\tau )dxdyd\tau 
\end{equation}
where D is the surface on which each point is situated at a distance $Z(P'(x,y,z),t,t')$  from the point A. The probability of nucleation in this domain within the period $t'$ to $t'+dt'$ is equal to

\begin{equation}
P_Adt'=\int _D \int Z(P',\tau ,t')v(P',\tau )J(t')dxdy dt'
\end{equation}
where $J(t')$ is nucleation rate per unit volume at time $t'$ . Then the probability that droplets formed at a randomly selected point during the period of time between $t'=0$ and $t'=t$ reach the point A is
\begin{equation}
P_A=\int _0^\tau \int _D \int Z(P',\tau ,t')v(P',\tau )J(t')dxdydt'
\end{equation}
The point A transforms to a stable phase from the metastable phase only if the point has not  transformed previously. Thus, the rate of  transformed is
\begin{equation}
\label{eq:co9}
df(\tau )=-dg(\tau )=g(\tau )P_A 
\end{equation}
The solution of Eq.(\ref{eq:co9}) is given by
\begin{equation}
\label{eq:co10}
\ln\frac{g(t)}{g(0)} = -\int _0^t d\tau \int_0^\tau \int _D \int Z(P',\tau ,t')v(P',t)J(t')dxdydt'
\end{equation}
Since the integral of the right hand side of Eq.(\ref{eq:co10}) has a finite value, we can change the sequence of integration. Then
\begin{equation}
\ln\frac{g(t)}{g(0)} = -\int _0^t J(t')dt' \int_{t'}^t  \int _D \int Z(P',\tau ,t')v(P',t)dxdyd\tau 
\end{equation}
The volume at a time $t$ of a droplet which formed at a time $t'$ is
\begin{equation}
v(t,t')=\int_{t'}^t\int _D \int Z(P',\tau ,t')v(P',t)dxdyd\tau 
\end{equation}
At $t=0$, $g(0)=1$ , we obtain the well known  Kolmogorov-Johnson-Mehl-Avrami type formula
\begin{equation}
f(t)=1-\exp\left[-\int_0^t J(t')V(t,t')dt \right]
\end{equation}


\begin{thebibliography}{99}
\bibitem{av}
M. Avrami:J.Chem.Phys, {\bf 7}, 1103(1939)
\bibitem{kol}
A.N. Kolmogorov: Izv. Akad. Nauk SSSR, Ser. Matem., No. {\bf 3}, 355(1937),
\bibitem{jm}
W.A. Johnson and R.F. Mehl Trans. AIME, {\bf 135}, 416(1939)
\bibitem{chr}
A.A. Chernov: "Modern Crystallography III Crystal Growth", Springer-Verlag,(1984)
\bibitem{bs}
K. Binder and D.Stauffer: Adv. Phys., {\bf 25}, 343(1976)
\bibitem{bs1}
K. Binder, D. Stuaffer and H. M${\rm \ddot{u}}$ller-Krumbhaar; Phys. Rev. B, {\bf 12}, 5261(1975)
\bibitem{bd}
R.Becker and W. D${\rm \ddot{o}}$ring: Ann. Phys., {\bf 24}, 719(1935)
\bibitem{couran}
R. Courant and F. John: Introduction to Calculus and Analysis I, (Springer-Verlag. New York, 1989)
\bibitem{kds}
I. Kanne-Dannetschenk and D.Stauffer: J.Aerosol.Sci., {\bf 12}, 105(1981)
\bibitem{sa}
Y. Saito, M. Honjo, Konishi and A. Kitada: J. Phys. Soc. Japan, {\bf 69}, 3304(2000)
\end{thebibliography}
\end{document}